# Exceeding Computational Complexity – Trial-and-Error, Dynamic Action and Intelligence[1]

Chuyu Xiong

Independent Researcher

*chuyux99@gmail.com

**ABSTRACT**

Computational complexity is a core theory of computer science, which dictates the degree of difficulty of computation. There are many problems with high complexity that we have to deal, which is especially true for AI. This raises a big question: Is there a better way to deal with these highly complex problems other than bounded by computational complexity? We believe that ideas and methods from intelligence science can be applied to these problems and help us to exceed computational complexity. In this paper, we try to clarify concepts, and we propose definitions such as unparticularized computing, particularized computing, computing agents, and dynamic search. We also propose and discuss a framework, i.e., trial-and-error + dynamic search. Number Partition Problem is a well-known NP-complete problem, and we use this problem as an example to illustrate the ideas discussed.

## 1   Introduction

Computing technology and theory are developing rapidly, however, the ultimate concern remains: How can we expand our practical computing power to engineering and scientific problems, including those in artificial intelligence?

One focus of these problems is computational complexity. Computational complexity is a central part of computer science, which studies how much time and how much storage space are required to compute a problem. Usually, computational problems have scale, such as the size of the matrix, the number of bits of integers, the environmental complexity of autonomous driving, the accuracy of visual recognition, and so on. The computational complexity grows with the scale of the problem, and may grow relatively slowly, generally expressed by polynomial growth, or it may grow rapidly, generally expressed by exponential growth. Problems with exponentially increasing complexity are commonly considered to be very difficult. The most typical example is the factorization of large integers. This is an exponentially increasing, extremely difficult computational problem, which turn out to be the theoretical basis for cryptography.

Problems with high computational complexity are undoubtedly very difficult, however, we cannot ignore these problems, we have to face them. Artificial intelligence is even more inseparable from these problems of high complexity. In almost all aspects of artificial intelligence, it is inevitable to encounter these problems. For example, face recognition is a NP-complete problem [1]. That means that when we deal with these problems, the requirement for computing time and storage is very high. So, we hope to adopt intelligent methods to reduce the requirement of time and storage. Thus, artificial intelligence is highly entangled with computation complexity: we need to exceed computational complexity for problems in AI, but in turn, we need intelligent methods to exceed complexity. Such entanglement will push us more and more to intelligence. This situation could be contrasted in parallel to life that develops and improvs the intelligence while dealing with difficulties.

---

[1] Thanks for my wife's consistent support

But we have to first clarify the meaning. Computational complexity is a solid and rigorous theory. The theoretical limitation set by the theory cannot be violated. However, we need to know that the limit set by the theory is for a whole set of instances of problems, not for a single instance. Yet, we often only need to solve a particular instance, not the whole set. We then ask: for a particular instance, is there a solution that is much better than the generic solution for the whole class? And, how to find a good particular solution for the particular instance? If we can do so, then, we can achieve a much better solution for a particular instance than following the generic solution. This is what we mean when we say exceeding computational complexity.

Exceeding computational complexity, no doubt is very difficult. In fact, whether or not it is possible to do so is a big question. At present, there is no solid theory to support it, and certainly no definite theory to completely deny it. Therefore, in this article, we try to make a preliminary discussion on this. We tried to clear out a discussion framework to facilitate further exploration. In fact, what motivates us to go in this direction is: when we were studying the number partition problem [2], we found that there are some cases that look very difficult, but are actually easy to solve, and yet there are some cases that are indeed true hard (a very twisted and invisible fence [3]). When we face these puzzles and think again and again, we gradually find that intelligence is looming in it.

In this paper, we propose the main idea: we should do "particularized computation", not "unparticularized computation", for those problems with high computational complexity. And, a computing agent with intelligence and subjectivity inside can find particularized computing for a particular instance. We propose a framework, namely trial-and-error + dynamic action, for such a computing agent. We demonstrate this idea with the number partition problem, which is one of the famous Karp's 25 NP-complete problems.

## 2   Limit of Computation – Computational Complexity

Any computation will require resources, be storage spaces or processor cycles, that are demanded by the computational complexity of this computation. This can even be seen as physical barrier: without the required resources available, it is physically impossible to execute the computation. In fact, as well known, modern cryptography is built on such physical requirement created by computational complexity.

However, there are different ways to deal with the limit of computation, and different ways will require different resources. We can consider a very simple example, $4567 \times 2341$. If the ordinary multiplication rules are used, 16 single-digit multiplications and 12 single-digit additions are required. Such a computation is good for this particular case, and is good for any other case . But what if our problem is $4567 \times 2300$? We can of course also use the above way to do, so we need to do 16 multiplications and 12 additions. However, obviously we can take a more compact approach, requiring only 8 multiplications and 4 additions to complete the computation. This is because we take full advantage of the particular properties of this particular problem, the 0 in 2300, so we can use much less resources.

From this simple example, we can see that the two ways to deal with a computational task. One is to use a same definite fixed program to all cases; the other is to use a particular method suitable for the particular case if it is possible. Just as the example shows, the latter way requires much less resources.

Let's look at a more realistic case. There is an integer array: $\Omega = \{63,48,932,266,671,47,110,82,39\}$, we want to divide this array into two arrays so that the sums of two arrays are equal. There is a method that can be applied to any array. It is the exhaustive search, that is, to go through all possible partitions. The length of this array is 9, so exhaustive search requires $2^8$ searches, which is quite computationally expensive (at least for manual way). However, for this particular case, we can try to solve it by a particular way. We are going to solve it by trying and adjusting. First, we divide it into two arrays arbitrarily: $\Omega_1 = \{63,48,932,266\}$, $\Omega_2 = \{671,47,110,82,39\}$. It is easy to see that the sum of first is bigger. Then, let's make some adjustments to make the sum of first smaller and the second bigger, like this: $\Omega_1 = \{63,48,932,110\}$, $\Omega_2 = \{671,47,82,39,266\}$. Now, the first is still bigger. However, this time, we can see it more clearly, and know how to do the final adjustment. We get: $\Omega_1 = \{63,48,932,47,39\}$, $\Omega_2 = \{671,82,266,110\}$. The sum of the two arrays is now equal. This way, the amount of computation is much smaller.

That is to say, when we have a particular case, if we try to find various particular relationships in the particular case, and utilize these relationships as much as possible, we could compute with much less resources. Therefore, the question naturally arises, does this approach indeed save resources? This is not an easy question. We should first make some definition carefully.

**Definition 2.1 (Unparticularized computation and particularized computation)**
*Suppose a computation problem, whose all instances form a set $W$, and we denote a computing program $C$ acts on the particular instance $w \in W$ as $C(w)$. If there is a fixed and definite computing program $C$, for any $w \in W$, $C(w)$ can get the correct result, we say that the computing program $C$ is a unparticularized program covering $W$. If for a particular instance $w$, there is a computing program $C_w$, so that $C_w(w)$ can get the correct result, and can do so with as few resources as possible, we say that $C_w$ is a particularized computing program for the instance $w$; however, note, for $u \in W$, $u \neq w$, $C_w(u)$ may not be able to get the result (not stopping or crash), or the result may be incorrect, or the resource consumption may be much larger.*

In the terms of this definition, we can see, in the above example of number partition problem, the method of exhaustive searching is unparticularized computing, since it works for all cases, the method of trying and adjusting is particularized computing since it works only for the particular case, and not for other cases.

When faced with a computation problem, the mainstream effort so far has been to try to find unparticularized computing programs. This is reasonable, because with unparticularized computing thinking becomes simple. For any instance, only need to "plug in" the instance into the computing program, and always obtain correct result, no any other care is required. Simplified thinking and always correct are what people desire. For easier problems, or with plenty of resources available, this approach is perfectly reasonable. However, such approach is no longer appropriate when dealing with problems of high complexity. Just as the above examples show, it is more reasonable to explore the particular properties of the particular instance, and to take full advantages of these properties. These particular properties could be like symmetry, weaknesses, patterns, etc., and they can be used to reduce the consumption of resources. That is to say, for problems with high complexity, we should pursue particularized computing, rather than unparticularized computing.

OK, for a particular instance, use particularized computing, instead of unparticularized computing. Sounds great. But where the particularized computing program comes from? For unparticularized computing, we know how to do. It is in this way: human programmers work hard to get a program that is working for all instances. We understand this well and are very familiar with. Whole industry is doing this for decades. However, we do not know how to do particularized computing, it is very unfamiliar to us. Should we develop a program manually for each particular instance? This simply is not practical. Or, should we have a finder program that can help us to find the particularized computing program for a particular instance? Looks great. But such a finder itself will consume resources. So, question comes: can the overall resource consumption be reduced? Thus, we need to make some definitions about resources.

**Definition 2.2 (Resources required by unparticularized computation)**
*Assuming a computation problem, all instances form a set $W$, if $C$ is a unparticularized computation program covering $W$, for each particular instance $w \in W$, the resources $C(w)$ consumes is $Z_w$, then the resources required by $C$ is: $Z = max\{Z_w | w \in W\}$. If there are more than two unparticularized computing programs covering $W$, the smaller of the 2 resources required for the 2 computing programs is taken as the resources required by unparticularized computing program covering $W$.*

For particularized computation, we have the following proposition.

**Proposition 2.3 (Resources required by particularized computation)**
*Assuming a computation problem, all instances form a set $W$, and the resources required by unparticularized computing covering $W$ are $Z$. Suppose we use a definite and fixed finder program $S$ to find the particularized computing program for a particular instance, that is, for any $w \in W$, $S(w)$ can get the particularized program $C_w$ for the instance $w$, then the upper bound of the resources consumed by $S(w)$ and $C_w(w)$ will be equal to $Z$.*

**Proof:** $S$ is a definite and fixed finder program, so for any instance $w \in W$, we first do $S(w)$ to get $C_w$, and then do $C_w(w)$ to compute the instance. Thus, this procedure forms an unparticularized computing program covering $W$, we denote it as $C$. Therefore, the resources required by $C$ are $Z$. ∎

This proposition tells us: it is not good to use a definite and fixed finder to find a unparticularized computing, since by this way, no resources could be saved. According to this proposition, the resources required by unparticularized computation is a standard standing well, which could not be easily overpass. Thus, we define the resources required for computation as below.

**Definition 2.4 (Resources required by computation)**
*Assuming a computing problem, all instances form a set $W$, the resources required by computation for this problem are the resources required by unparticularized computing covering $W$.*

That is, for an instance $w \in W$, the computation requires resources $Z$, which is defined by unparticularized computing. $Z$ is the limit dictated by computational complexity. But if we know a particularized computing program for $w$, then we could do computation with resources much less than $Z$. This is exceeding computational complexity. It is great if we can do so. However, it needs us to know the particularized program for $w$. If we do not know such program in advance, can we still exceed computational complexity? If so, how? Certainly, not by a definite and fixed finder program, as Proposition 2.3 tells. Intelligence will be essential.

## 3  Computing Agent and Trial-and-error + Dynamic Action

For a given particular instance, the question is how to find the particularized computing program that requires much less resources. We cannot manually find such program, neither use a definite and fixed finder. Here are some possibilities to have particularized program. A) We just have it. B) We have memory of a lot of such programs and have a looking table to locate it. C) We will interact with the particular instance and then get the particularized program from the interaction, and such process only consumes an order lower of resources.

A) is like an oracle. This is very interesting. It shows a strong connection between particularized computing and non-deterministic Turing machine. But we will not consider it now. B) does not work as well. It requires to remember all instances. Normally the number of instances is very huge, to remember will need a lot of resources. The possibility C) means the computing entity to do particularized computation has some ability. In fact, a very strong ability: it can explore the situation, make judgment and utilize possibilities just popup.

We would like to call such a computing entity as a computing agent, and this agent has intelligence and subjectivity inside. Wang Pei is a researcher and advocate of AGI, according to the definition of intelligence he advocates: intelligence is the ability to make the best adaptation under the circumstance of limited resources [4]. Obviously, his definition of intelligence is in the same direction as that we call the ability of computing agent as intelligence. In fact, this definition of intelligence that given by Wang Pei has a positive effect on us. Based on these considerations, we have the following definitions.

**Definition 3.1 (Intelligence of Computing Agent)**

*Suppose there is a computational problem, all instances form a set $W$, and then suppose that the resources necessary for unparticularized computing covering $W$ are $Z$. Now there is a computational agent $A$ to deal with $W$, if for instance $w \in W$, $A$ can have a particularized program $C$, and $C(w)$ can be done with resources one order of magnitude lower than $Z$ ($O(log(Z))$), then we say that $A$ can intelligently compute $w$. If $V \subset W$ is the subset of all elements that $A$ can intelligently compute, then the intelligence of $A$ is measured by the quantity: $q(A) = |V|/|W|$.*

That is to say, a computing agent with intelligence greater than zero can break through the barriers of computational complexity. However, a fundamental question is: Does such a computing agent really exist? To the best of our knowledge, there is currently no theory discussing whether such a computing agent exists. Furthermore, there is no theory that tells us how to build an intelligent computing agent. These problems are not only major problems in computational theory but also major theoretical problems in artificial intelligence, which require further research. Here, to make an attempt, we propose this idea: if appropriate trial-and-error procedures and dynamic action are adopted, there is hope to form an intelligent computing agent.

In our common sense, trial-and-error is very reasonable method, in fact, we often use it unconsciously. In the theory of computation, Gold, Putnam, Kugel et al. insisted on using the trial-and-error method to deal with the problem of computability [5]. Kugel called such a trial-and-error computing program a Gold-Putnam machine. They thus developed a trial-and-error procedure for dealing with the more difficult computability problems, as well as trying to deal with the non-computable ones.

What exactly is trial-and-error doing? Why can it be successful? Trial-and-error is actually based on the following facts: 1) admit that we have an unknown, but this unknown can be obtained through effort; 2) some transformation can be used to transfer the unknown to a definite search space; 3) trial-and-error efforts is a cycle: to get feedback from trial, to seek better by feedback, and seeking is to move from one point to the next point in the search space; 4) can reach (or get close to) the unknown in the search space. Trial-and-error is a very effective mechanism and often an indispensable tool for solving problems.

Back to our problem of finding particularized computation. Assuming all instances of the problem form a set $W$, given an instance $w \in W$, we want to find the particularized computing program $C_w$ for $w$. So, here the unknown is $C_w$. As discussed earlier, we want to transform the unknown into some search space. There can be many kinds of search spaces. We are here to make certain restrictions. We restrict the search space to a Boolean vector space, that is, the search space is $B^N$. Such restrictions are of course limited. However, if $N$ is large enough, the Boolean vector space can actually cover any parameter space, and it is very common to use the parameter space to regulate the computation (as shown by the non-deterministic Turing machine, see Cook [6]). Therefore, it is reasonable to choose the search space as the $N$-dim Boolean vector space. We can change to a different search space later if necessary. Thus, the search is to obtain the correct parameter vector $p \in B^N$, and then the parameter vector $p$ will bring the particularized program $C_w$ for $w$ to us. Since $C_w$ depends on the parameter vector, we can write it as $C(w, p)$.

With the search space in hand, let's consider a trial-and-error procedure. We need to have these components for such procedure. First, a trial-and-error program $T(w, p), w \in W, p \in B^N$, if the given parameter vector $p$ is correct, $T(w, p) = 1$, otherwise $T(w, p) = 0$. This is the major feedback. But, $T$ also feeds back other information $t$. Second, a search program $S(w, p, t), w \in W, p \in B^N$, $S$ will yield $p_n$, which is the next point in the search space for trial-and-error. $S$ also produces some other information for trial-and-error use. Third, a computation program $C(w, p), w \in W, p \in B^N$, if parameter vector $p$ is correct, this program will be the particularized program. With these components, the trial-and-error procedure is as follows:
   1) Initially set the parameter $p = p_0$ and start a trial-and-error cycle.
   2) Trial-and-error cycle: Run trial-and-error $(c, t) \leftarrow T(w, p)$, where $c$ is the testing result, and $t$ are all other feedback information. If testing result is 1, the parameter vector is correct, exit the cycle. If testing result is 0, continue.
   3) Search $S(w, p, t)$, $S$ generates $p_n$, which is the parameter vector used for the next trial-and-error.
   4) The trial-and-error cycle continues until the correct parameter $p$ is obtained, or an error is reported.
   5) If the correct parameter vector is obtained, the particularized program $C(w, p)$ is also obtained.

Note that in the trial-and-error procedure, the most important component is $S(w, p, t)$, which will generate next parameter vector for trial. $S$ could use the dumbest way, exhaustive search, to search every possible point. However, this is not what we expected. For exhaustive search, $2^N$ resources must be used. We expect to use much less resources. So, we need to have a much better $S$, dynamic search, which has the ability to intelligently use the feedback information from trial-and-error. In order to use the feedback information intelligently, dynamic search needs to have its subjectivity. We discussed subjectivity and dynamic action of machine in detail in [7]. Now we can define intelligent search.

**Definition 3.2 (Intelligent Search)**
*Assuming that the search space in the trial-and-error procedure is $P = B^N$, and the dynamic search is $S(w, p, t)$, if for a given $w$, for any initial parameter $p_0$, $S$ can reach the correct parameter vector $p$ for $w$ by using only $O(N)$ resources, we say that $S$ can do intelligent search for $w$.*

A computing agent with intelligent search is really intelligent.

**Proposition 3.3 (Trial-and-error + Dynamic Action)**

*Suppose there is a computational problem with scale $N$, and all instances forms a set $W$, and the resources required for unparticularized computation covering $W$ are $O(2^N)$. Suppose that the computing agent $A$ consists of a program $C(w, p)$ with parameters and trial-and-error procedure + dynamic search $S$. For an instance $w \in W$, if $S$ can do intelligent search for $w$, and the program $C(w, p)$ only needs $O(N)$ resources, then the computational agent $A$ has intelligence as defined in 3.1.*

It is easy to see that this proposition is true: If both $S$ and $C(w, p)$ only need resources of $O(N)$, so the set $V \subset W$ specified in Definition 3.1 is not empty, thus, the computational agent $A$ has intelligence. So, $A$ can exceed computational complexity.

Intelligent search and computing agent are great. But, how can we get them? This is a big issue and requires a lot of further work. In next section, we will use number partition problem as one example to shed some light on it.

## 4   Number Partition Problem

Number partition problem is a very famous and important problem. We now use this question as an example and apply the ideas discussed earlier in the hope that it will help us to see things better. Number partition problem can be explained in a short sentence: given a set of natural numbers $\Omega$, ask whether $\Omega$ can be divided into two subsets $\Omega_1$ and $\Omega_2$ such that the sum of the numbers in $\Omega_1$ equals the sum of the numbers in $\Omega_2$? This problem is a typical example of P vs. NP: it is easy to verify solution but hard to find.

Let's describe the problem in more detail. Now let the array length be $N$, we will consider the set $\Omega$ of length $N$, whose members are all natural numbers, that is, $\Omega \in I^N$, $I$ is the set of natural numbers. Then, we consider a $N$-dim Boolean vector $p = (p_1, p_2, \ldots, p_N), p_j \in B, p \in B^N$ and for a set $\Omega = \{\omega_1, \omega_2, \ldots, \omega_N\}$ and a Boolean vector $p$, we define a quantity:

$$< p, \Omega > = \sum_{j=1}^{N} q_j \omega_j \quad \text{if } p_j=1, q_j=1; \text{ if } p_j=0, q_j=-1 \tag{1}$$

It is easy to see that the meaning of the quantity in (1) is: partition the set into two subsets, and the quantity is the difference between the sum of the two parts. Clearly, the parameter vector $p$ tells how to partition $\Omega$ into two subsets, so, we call $p$ a partition vector. That is to say, this quantity $< p, \Omega >$ is actually a test whether the partition vector $p$ can equally partition $\Omega$, and this quantity also gives the feedback about how far away the partition is from equal partition. Let's define another function: $\varphi(\Omega, p)$:

$$\varphi(\Omega, p) = \begin{cases} 1 & \text{if } < p, \Omega > = 0 \\ 0 & \text{if } < p, \Omega > \neq 0 \end{cases} \tag{2}$$

$\varphi(\Omega, p)$ is a boolean function with parameters, $\varphi(\Omega, p): I^N \times B^N \to B$, that is, if the partition vector $p$ equally partition $\Omega$, the function value is 1, otherwise it is 0. So, this function is a trial-and-error function, using $p$ as the parameter vector. We defined an operator $\odot$ in [2], which means "trying over". Using function $\varphi(\Omega, p)$ and this operator $\odot$, we can define the partition function:

$$Par_N(\Omega): I^N \to B, \quad Par_N(\Omega) = \varphi(\Omega, p) \odot P \tag{3}$$

```
Pseudo Code: Trial-and-Error + Exhaustive Search
Setting: $p_0$, $\Omega$
Initial: $N = 0$, $p = p_0$
While $N < 2^N$
   Continue to trial and error: $t \leftarrow <p, \Omega>$
   If $t = 0$ then
       Stop, output "Computing successful, $Par_N(\Omega) = 1$, Partition vector: "$p$
   Else
       Continue search: $p_n \leftarrow S$, $p \leftarrow p_n$, $N \leftarrow N + 1$
   EndIf
EndWhile
Stop, output "Computing successful, $Par_N(\Omega) = 0$"
```

Here, $P$ is the space formed by all partition vector (here $P = B^N$). The function means: apply all $p \in P$ to $\varphi(\Omega, p)$ to try it out. If any value in the test result is equal to 1, then the value of $Par_N(\Omega)$ is 1, otherwise the value of $Par_N(\Omega)$ is 0. That is, the function value of $Par_N$ is defined by trial-and-error. This clearly tells us that the definition of the problem of number partition is defined by "trial-and-error + exhaustive search". So, quite naturally, we can directly translate this definition into the computation. See pseudo code "Trial-and-Error + Exhaustive Search".

Obviously, the previous trial-and-error procedure + exhaustive search is just the program expressing $Par_N(\Omega) = \varphi(\Omega, p) \odot P$. In fact, the program and equation are exactly the same thing. Therefore, using trial-

```
Pseudo Code: Trial-and-Error + Dynamic Search
Setting: $p_0$, $N_{max}$, $\Omega$
Initial: $N = 0$, $p = p_0$
While $N < N_{max}$
   Continue to trial and error: $t \leftarrow <p, \Omega>$
   If $t = 0$ then
       Stop, output "Computing successful, $Par_N(\Omega) = 1$, Partition vector: " $p$
   Else
       Continue to search: $(p_n, i) \leftarrow S(t, p, \Omega)$
       If $i = 0$ then
               Stop, output "Computing successful, $Par_N(\Omega) = 0$"
       Else $i = 1$
               $p \leftarrow p_n$, $N \leftarrow N + 1$
       Else
               Stop, output "Computing failed, search stops"
       EndIf
   EndIf
EndWhile
Stop, output "Computing failed, out of range"
```

and-error + search to compute $Par_N$ is very natural. In the program, the search $S$ is an exhaustive search, i.e., $S$ walks through the entire $P$. There are many ways to implement an exhaustive search, as long as $S$ can traverse the entire $P$.

We would emphasize that the above computation is unparticularized computation, which is applicable to any instance and always get correct result. It is worth noting that the resources required by this computation are $O(2^N)$. Now we have unparticularized computation for number partition problem. How can we have particularized computation? As we discussed in Section 3, we need to transform to a search space. For number partition problem, it is relatively easy, since the definition of partition function already contains the search space. But we need to change the exhaustive search to dynamic search. Then, the computation program is shown in pseudo code "Trial-and-Error + Dynamic Search".

Note, in trial-and-error + dynamic search, the program is different than in trial-and-error + exhaustive search in several places. First, dynamic search does not guarantee the success of search. The program has 4 exits. Two exists are for successful, corresponding to $Par_N(\Omega)$ equal to 1 or 0. The other two exist are for failures, one failure is because the number of searches exceeds the preset value and is forced to stop, and the other failure is because the dynamic search thinks that the search can no longer be continued. Thus, the program does not always success. When the program successes, the value of $Par_N(\Omega)$ is given, otherwise the value of $Par_N(\Omega)$ cannot be given. It should also be noted that when the value of $Par_N(\Omega)$ is given as 1, the partition vector is also given, and both value and partition vector are guaranteed to be correct. However, when the value of $Par_N(\Omega)$ is 0, it may be right or wrong, because the search is not an exhaustive search, but a dynamic search with intelligence and subjectivity, which could be wrong. In short, the computing agent does a particularized computing for an instance (the given number array). There 3 kind results: 1) computation is done, and the computation result is correct, 2) computation is done, but the result is wrong, 3) computation fails, and there are 2 kind failures, one is that the search stops, and the other is out of range.

The most important part in the program is the dynamic search $S(t,p,\Omega)$. It is a computing agent with intelligence and subjectivity. It will take a look on its inputs: $t,p,\Omega$, where $t$ is the feedback information from the testing program, and $p$ is partition vector currently using, and $\Omega$ is the number array that currently doing. $S$ will intelligently use the information to decide what is the partition vector will be used to try next time. $S$ will not act in a presetting way, but it will be able to tell the current situation, explore the possibilities, and try to give the best guess for next partition vector, and save the resources to be used. $S$ should be able to learn from the situation and its mistakes as well. So, $S$ is definitely not a traditional search program. But the question is: can we make such dynamic search with intelligence and subjectivity inside?

For the number partition problem, for any given number array $\Omega$ that can be equally partitioned, there is always a search that very quickly get the partition vector of $\Omega$, because we can set the initial vector $p_0$ as the partition vector. However, such searches are mundane, trivial, and not the kind of dynamic search we really want. The dynamic search we hope is like this: starting from any initial vector $p_0$, it will be able to reach correct results by only using much less resources. We propose the following conjecture.

**Conjecture 4.1 (Dynamic search exists for number partition problem)**
*For number partition problem, there is a dynamic search $S(t,p,\Omega)$ with such properties: for most integer array $\Omega \in I^N$, and for any initial partition vector $p_0$, $S$ will be able to reach correct results by only using $O(N)$ resources. The meaning of reaching correct*

*results is: the trial-and-error + dynamic search will make computation done (not exit abnormally), and correctly give the value* $Par_N(\Omega)$.

If we can indeed demonstrate such a dynamic search, it will be a major break through in computational theory that can be used in many engineering application areas. Not only that, finding such dynamic search will also be a solid progress of intelligence science. However, we point out: even we find such dynamic search, the fact that number partition problem is a NP-complete problem will not change, i.e., there are some $\Omega \in I^N$ so that the trial-and-error + $S$ will not be able to compute $Par_N(\Omega)$ by only using $O(N)$ resources.

## 5   Remarks

We would like to emphasize our major points again. Computational complexity sets a limit: for a computation, there must be enough resources available for it, if there are no enough resources, the computation could not be achieved. But this limit is for all instances of the problem. For a particular instance, it is possible to use particularized program to achieve computation that will only requires much less resources. This is exceeding computational complexity. However, in order to have particularized program, we need computing agent with intelligence and subjectivity inside.

But we have to say, it is very controversial. So far no one have made such an agent. And there is no theory to fully support such agent yet. But we strongly believe that such computing agent does exist, and there is huge demand for it since there are many hard problems are waiting for such agent. Thus, we try to clear mist around this issue and establish some solid ground for further discussions. We clean up some crucial concepts such as unparticularized and particularized computing, trial-and-error, dynamic search, etc. Number partition problem, due to its nature, can serve as one good example for these. For this problem, we conjecture the existence of computing agent. Note, number partition problem is one NP-complete problem. If we can find a computing agent for it, the computing agent then can be used for many other hard problems. The direction is clear now. We will continue research along this path and hope to reach a concrete computing agent.

Here, we would like to mention the close relationship between particularized computing and non-deterministic Turning machine [6]. We think this relationship is a major problem in computational theory and in artificial intelligence. AI and exceeding computational complexity are very deeply entangled together. This is a good thing actually. Such entanglement will strongly help the development of artificial intelligence.

We can see the relationship between particularized computing and intelligence from another view. If there is no difficult problem, all computations can be done by unparticularized computing, there will be no need for intelligence. For highly complex computational problems, only a particularized computing has hope to do it. Yet, only intelligence can make particularized computing possible. This is the reason that intelligence exists and must exist. We can measure intelligence by how well it can create particularized computing. If we see a computing agent can create nice particularized computing in a difficult situation, we see intelligence inside it.

**References**

[1] A. Yang, J. Wright, Y. Ma and S. Sastry, "Feature Selection in Face Recognition: A Sparse Representation Perspective," 2007.

[2] C. Xiong, "Sampling and Complexity of Partition Function," *arxiv.org,* 2021.

[3] S. Aaronson, "P =? NP," 2011.

[4] P. Wang, "Three fundamental misconceptions of artificial intelligence," vol. 19, no. 3, pp. 249-268, 2007.

[5] P. Kugel, "Thinking may be more than computing".

[6] S. Cook, "THE P VERSUS NP PROBLEM," 2000.

[7] C. Xiong, "Some Discussions on Subjectivity of Machine and its Function - Contributions to ICIS 2020," in *International Conference of Intelligence Science 2020*, West Bengal, India, 2020.